# Phase diagram of (Li$_{1-x}$Fe$_x$)OHFeSe: a bridge between iron selenide and arsenide superconductors

Xiaoli Dong,[*,†] Huaxue Zhou,[†,‡] Huaixin Yang,[†] Jie Yuan,[†] Kui Jin,[†] Fang Zhou,[†] Dongna Yuan,[†] Linlin Wei,[†] Jianqi Li,[†] Xinqiang Wang,[‡] Guangming Zhang,[§] and Zhongxian Zhao[*,†]

[†] Beijing National Laboratory for Condensed Matter Physics, Institute of Physics, Chinese Academy of Science, Beijing 100190, China

[‡] College of Physics, Chongqing University, Chongqing 401331, China

[§] Department of Physics, Tsinghua University, Beijing 100084, China

**ABSTRACT:** Previous experimental results have shown important differences between iron selenide and arsenide superconductors, which seem to suggest that the high temperature superconductivity in these two subgroups of iron-based family may arise from different electronic ground states. Here, we report the complete phase diagram of a newly synthesized superconducting (SC) system (Li$_{1-x}$Fe$_x$)OHFeSe with a similar structure to FeAs-based superconductors. In the non-SC samples, an antiferromagnetic (AFM) spin-density-wave (SDW) transition occurs at ~127 K. This is the first example to demonstrate such an SDW phase in FeSe-based superconductor system. Transmission electron microscopy (TEM) shows that a well-known √5×√5 iron vacancy ordered state, resulting in an AFM order at ~ 500K in A$_y$Fe$_{2-x}$Se$_2$ (A = metal ions) superconductor system, is *absent* in both non-SC and SC samples, but a unique superstructure with a modulation wave vector $q$=1/2(1, 1, 0), identical to that seen in SC phase of K$_y$Fe$_{2-x}$Se$_2$, is dominant in the optimal SC sample (with an SC transition temperature $T_c$ = 40 K). Hence, we conclude that the high-$T_c$ superconductivity in (Li$_{1-x}$Fe$_x$)OHFeSe stems from the similarly weak AFM fluctuations as FeAs-based superconductors, suggesting a universal physical picture for both iron selenide and arsenide superconductors.

The iron-based superconductors, containing similar blocks of either FeSe or FeAs layers, have triggered extensive attention due to their rich structural and physical properties.[1-5] Previous reports have revealed that, though they share a similar structural skeleton, these two subgroups of iron-based family exhibit quite different and puzzling features of their physical properties. The simplest binary β-FeSe material with the structural symmetry of P4/nmm shows a superconductivity at $T_c$ ~ 8.5 K.[6,7] When alkali or alkali-earth metal ions are intercalated into the adjacent FeSe layers, the value of $T_c$ of the derived materials A$_y$Fe$_{2-x}$Se$_2$ [5,8-18] is increased up to 46 K, but its structural symmetry is changed to I4/mmm. In contrast to the FeAs-based superconductors, the materials A$_y$Fe$_{2-x}$Se$_2$ always manifest themselves as a mixture of high-$T_c$ superconductivity, strong AFM phases and various cation/iron vacancy ordered phases.[13,19-28] For example, the superconducting phase always coexists with the insulating phase (245) with a √5×√5 Fe vacancy ordered state, which displays a strong cluster AFM order at about 500K. Unlike a metallic background in iron arsenide materials, the electric resistivity of A$_y$Fe$_{2-x}$Se$_2$ shows a broad hump between about 70K and 300K, another indication of the mixed electronic states inside the materials. Because of the phase-separation in the FeSe-based superconductors,[5,13,24] a dome-like doping dependence of $T_c$ has not as commonly been observed so far as in the FeAs-based superconductors. Another notable difference is that the correlated AFM SDW order occurs in the FeAs-based parent compounds in the temperature range of 100-150 K[1-4,29], which is much lower than the FeSe-based counterparts. This is resulted from significantly different magnetic moments of irons in these two systems. So it would not be safe to assume a common SC mechanism to be shared by both subgroups of iron-based superconductors.

Recently a new superconductor (Li$_{0.8}$Fe$_{0.2}$)OHFeSe with $T_c$ about 40 K has been synthesized.[30-33] In this paper we report a complete phase diagram for (Li$_{1-x}$Fe$_x$)OHFeSe, which sheds a new light on the underlying physics in FeSe-based high-$T_c$ superconductors. We find that the physical properties of the non-SC samples of (Li$_{1-x}$Fe$_x$)OHFeSe are very similar to those of FeAs-based superconductors.[1-4,29] An emergent AFM SDW transition is observed at ~127 K in the absence of the well-known √5×√5 Fe vacancy ordered state associated with the AFM order in A$_y$Fe$_{2-x}$Se$_2$. In the optimal SC sample of (Li$_{1-x}$Fe$_x$)OHFeSe with $T_c$ = 40 K, one dominant superstructure with a unique modulation wave vector $q$=1/2(1, 1, 0) is observed by means of selected-area electron diffraction and high-resolution TEM. The same superstructure has also been reported for the SC phase of K$_{0.5}$Fe$_2$Se$_2$.[27] Therefore, the unusual (Li$_{1-x}$Fe$_x$)OHFeSe system bridges the gap between iron selenide and arsenide superconductors, and a universal physical picture of high-$T_c$ superconductivity is thus expected.

Hydrothermal syntheses of a series of (Li$_{1-x}$Fe$_x$)OHFeSe samples were carried out in stainless steel autoclaves of 50 ml capacity with Teflon liners.[33] 0.0075mol selenourea (Alfa Aesar, 99.97% purity), 0.0056-0.0075 mol Fe powder (Alfa Aesar, 99.998% purity), and 6 g LiOH·H$_2$O (Alfa Aesar, 99.996% purity) were mixed with 10 ml de-ionized water and loaded into the autoclave. The autoclaves were tightly sealed and heated in the temperature range of 120 - 180 °C for 3 - 4 days to obtain various samples (the details are shown in Table 1). The obtained products were washed by de-ionized water using Buchner Flask. Grains with size larger than 8 micrometers were collected and stored in liquid nitrogen due to their air sensitivity.[32] The structure of the polycrystalline samples was characterized by high-powder X-ray diffraction (XRD, an 18KW MXP18A-HF diffractometer with Cu–K$\alpha$ radiation). High-resolution TEM observations and microstructure analysis were performed on a Tecnai-F20 transmission electron microscope with a field emission gun operated at an accelerating voltage of 200 kV. Magnetic properties were determined by a SQUID magnetometer (Quantum Design MPMS XL-1). The superconducting transition of each sample was monitored down to 4.5 K under an external magnetic field of 1 Oe, while the SDW transition was measured under a field of 1000 Oe.

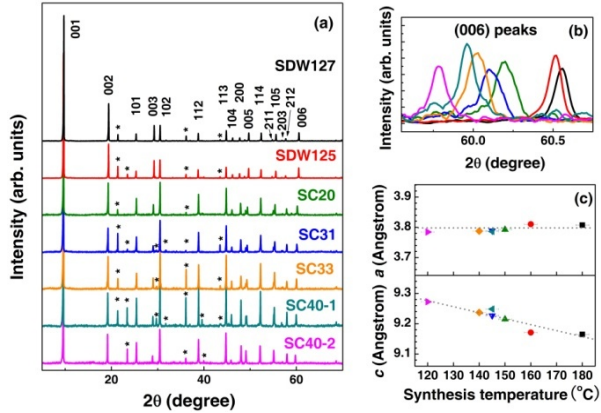

Figure 1. (a) Powder XRD patterns for the seven $(Li_{1-x}Fe_x)OHFeSe$ samples. Peaks of an unknown minor phase are indicated by stars. (b) Zoom-in (006) peaks. The position shift from right to left corresponds to the lattice expanding along $c$-axis. (c) The lattice constants of the $a$- and $c$-axes versus hydrothermal synthesis temperature for samples of SDW127, SDW125, SC20, SC31, SC33, SC40-1, and SC40-2 (from right to left).

It has been reported that,[30,32,33] unlike other iron selenide superconductors $A_yFe_{2-x}Se_2$, $(Li_{0.8}Fe_{0.2})OHFeSe$ is of the same crystal structure with the space group of P4/nmm as the quaternary iron arsenide superconductors. Figure 1a shows the powder XRD patterns of our seven $(Li_{1-x}Fe_x)OHFeSe$ samples, henceforth denoted by SC40-2, SC40-1, SC33, SC31, SC20, SDW125, and SDW127, respectively. All the diffraction peaks of each $(Li_{1-x}Fe_x)OHFeSe$ main phase can be well indexed with the previously reported tetragonal structure.[30,32,33] The least-squares refined unit cell dimensions, $a$'s and $c$'s, for all the samples are listed in Table 2. It is generally observed that the lattice constant $c$ tends to increase with the decrease of synthesis temperature (Figure 1b & 1c), while the lattice constant $a$ is reduced slightly (Figure 1c). This temperature dependence of $c$-axis parameter can be rationalized by the fact that lowing temperature is beneficial for the intercalation of the spacer layer $(Li_{1-x}Fe_x)OH$ of a higher $Fe_x/Li_{1-x}$ ratio in between the FeSe layers, so that the spacing of the FeSe layers is enlarged (note that $Fe^{2+}$ is bigger than $Li^+$). Although our ICP-AES (inductively coupled plasma atomic emission spectroscopy) results show that the total Fe/Li ratio also increases roughly with decreasing temperature of synthesis, so far it is difficult to accurately deduce the interstitial $x$ due to the variation of Fe vacancy in the FeSe layers.[32] Nevertheless, the lattice constant $c$, which is positively related to the doping level, can be accurately determined as a control parameter of the doping level in $(Li_{1-x}Fe_x)OHFeSe$. So we choose the lattice parameter of $c$-axis to plot the global phase diagram.

Table 1. Synthesis Conditions of $(Li_{1-x}Fe_x)OHFeSe$ Samples

| Sample | Fe(mol) | Temperature (°C) | Duration (Hours) |
|---|---|---|---|
| SDW127 | 0.0056 | 180 | 72 |
| SDW125 | 0.0056 | 160 | 72 |
| SC20 | 0.0056 | 150 | 72 |
| SC31 | 0.0056 | 145 | 96 |
| SC33 | 0.0075 | 140 | 72 |
| SC40-1 | 0.0056 | 145 | 72 |
| SC40-2 | 0.0056 | 120 | 96 |

Table 2. Unit Cell Parameters[*] and Volumes of $(Li_{1-x}Fe_x)OHFeSe$ Samples

| Sample | $a$ (Å) | $c$ (Å) | $V$ (Å$^3$) |
|---|---|---|---|
| SDW127 | 3.8073(5) | 9.1652(11) | 132.85 |
| SDW125 | 3.8104(4) | 9.1710(9) | 133.15 |
| SC20 | 3.7924(4) | 9.2152(8) | 132.54 |
| SC31 | 3.7891(2) | 9.2270(5) | 132.47 |
| SC33 | 3.7876(7) | 9.2373(14) | 132.52 |
| SC40-1 | 3.7860(4) | 9.2485(8) | 132.56 |
| SC40-2 | 3.7843(3) | 9.2729(7) | 132.80 |

[*] Well defined peaks with $2\theta$ smaller than ~ 60 degree are used to refine the cell parameters.

It can be seen from Figure 1a that there exist several peaks of an unknown minor phase, more or less, in the XRD patterns. The strongest peak is detected in sample SC40-1 with an integrated intensity of 14.7%. By a trial-and-error indexing procedure, we have searched out a bigger primitive tetragonal unit cell with refined lattice parameters $a_{mp}$ = 12.010(4) Å and $c_{mp}$ = 8.87(1) Å, on which the unknown peaks of sample SC40-1 can be indexed with acceptable deviations. The results of refinement are given in Table 3. The lattice dimension $c_{mp}$ is significantly smaller than those corresponding $c$-axis parameters of the main phases, therefore the minor phase can neither intergrow with $(Li_{1-x}Fe_x)OHFeSe$ nor be superconducting (see below). However, the lattice dimension $a_{mp} \approx \sqrt{10}a$, suggesting the minor phase to be some derivative from the basic tetragonal structure.

Table 3. Unit Cell Parameters and Indexing Results of the Minor Phase in Sample SC40-1

| $a_{mp}$ = 12.010(4) (Å) | $c_{mp}$ = 8.87(1) (Å) | $V_{mp}$ = 1279.07 (Å$^3$) | |
|---|---|---|---|
| h k l | $d_{calc}$ | $d_{obs}$ | $I_o$ (%)[1] |
| 1 0 2 | 4.160 | 4.159 | 11.3 |
| 3 1 0 | 3.798 | 3.800 | 14.5 |
| 4 0 0 | 3.002 | 3.005 | 7.1 |
| 3 3 0 [2] | 2.827 [2] | 2.816 [2] | 2.5 [2] |
| 4 0 2 | 2.486 | 2.487 | 14.7 |
| 5 1 1 | 2.276 | 2.275 | 5.1 |
| 4 1 3 | 2.075 | 2.075 | 7.2 |

[1] Integrated intensity relative to the main phase's (001) reflection.
[2] This very weak peak is excluded from the present refinement, though when included it can be indexed by (330) plane with $d_{calc}$ = 2.827 Å, leading to $a'_{mp}$ = 11.99(1) Å and $c'_{mp}$ = 8.89(4) Å.

Magnetic properties of all the seven samples have been investigated down to 4.5 K. Results shown in Figure 2a are the temperature dependence of their magnetic susceptibilities measured under 1 Oe after zero-field cooling. An SC transition at 40 K is observed in SC40-2, whose diamagnetic susceptibility is nearly -100% indicating a high sample quality. With shrinking $c$-axis lattice spacing as the doping decreased, the onset $T_c$ of $(Li_{1-x}Fe_x)OHFeSe$ changes gradually from 40 K (for SC40-2 and SC40-1), to 33 K (for SC33), 31 K (for SC31), and 20 K (for SC20). When the c-axis constant is smaller than about 9.2 Å, no superconductivity is detected. Instead, an emergent AFM SDW transition (Figure 2b) occurs at 125 K and 127 K for the samples SDW125 and SDW127, respectively. Based on our careful analysis, this SDW transition is not related to the minor phase detected by the XRD. The SDW transition temperatures are much lower than that of the AFM transition of $A_yFe_{2-x}Se_2$ (its Neel temperature $T_N \sim 500$ K).[5,13,19-21,24,25] By contrast, such a SDW transition at 100-150 K has been commonly seen in the FeAs-based parent compounds of REFeAsO (RE = rare earth element),[1-4] whose structural skeleton containing FeAs layers is of the same P4/nmm symmetry as the present $(Li_{1-x}Fe_x)OHFeSe$ samples. The similar SDW transition indicates that its magnetic coupling strength is comparable to the FeAs-based counterparts, in consistent with the fact that their $T_c$'s are similar. Therefore, this finding is of fundamental importance in the sense that this $(Li_{1-x}Fe_x)OHFeSe$ is so far the first found FeSe-based superconductor system associated with the presence of a weak AFM SDW order.

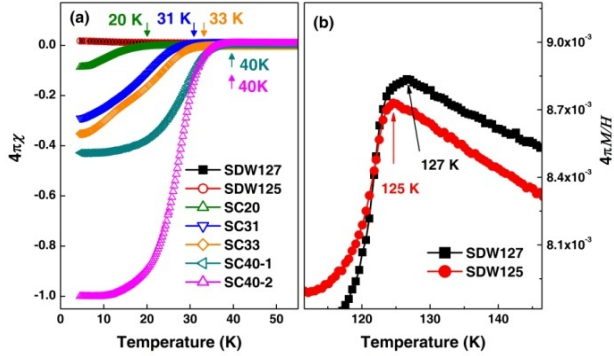

Figure 2. (a) Zero-field cooling magnetic susceptibilities of $(Li_{1-x}Fe_x)OHFeSe$ samples as functions of temperature. A nearly 100% diamagnetic shielding is achieved in the sample SC40-2. (b) Magnetization measured under an applied field of 1000 Oe. The AFM SDW transitions occur at 125 K and 127 K in SDW125 and SDW127, respectively.

The global phase diagram of $(Li_{1-x}Fe_x)OHFeSe$ is summarized in Figure 3a, by including the SDW and SC transition temperatures as functions of the lattice constant of c-axis. Obviously, the superconductivity is enhanced with expanding spacing of neighboring FeSe layers. It is noticed that the similarly positive $T_c$ dependences on the separation of FeSe layers have also been observed on different samples of $K_yFe_{2-x}Se_2$.[15,16] This phase diagram is very much similar to that of quaternary FeAs-based superconductors[1-4]: when the AFM SDW order is suppressed with increasing doping (i.e. with the increase in lattice spacing of c-axis in Figure 3a), a superconducting dome appears concomitantly with the optimal $T_c$ = 40 K.

In order to reveal the microstructure features in $(Li_{1-x}Fe_x)OHFeSe$ system, an extensive investigation has been performed at room temperature by means of selected-area electron diffraction and high-resolution TEM observation on non-superconducting (SDW127) and optimal superconducting (SC40-2) samples. Shown in Figure 3b and 3c are two typical electron diffraction patterns taken along the [001] zone axis. The main diffraction spots with a relatively strong intensity of both samples can be well indexed by the known tetragonal structure in consistence with the XRD results. Two striking features are revealed by our TEM observations. First, the superstructure of $\sqrt{5}\times\sqrt{5}$ Fe vacancies, being present in $A_yFe_{2-x}Se_2$ and resulting in the strong AFM order ($T_N \sim 500$ K), has not been detected in either SC or non-SC samples. Second, a superstructure is dominant in the optimal SC sample as illustrated in Figure 3c, in contrast to the non-SC sample (Figure 3b). The satellite spots, which generally are clearly visible in the $a^*$-$b^*$ plane of reciprocal space, can be characterized by a unique modulation wave vector $q = 1/2(1, 1, 0)$. Interestingly, a previous study[27] has also shown that the same superstructure dominates in the SC phase of $K_yFe_{2-x}Se_2$ with the chemical formula $K_{0.5}Fe_2Se_2$ and without the iron vacancy order. Such a unique superstructure, dominating simultaneously only in the SC phases of the two different FeSe-based compounds, may be associated with the intrinsic electronic states for superconductivity. At present stage, the origin of this superstructure is unknown. Whether and how it is microscopically correlated with spin and/or charge ordered states are very important issues worthy of further investigations.

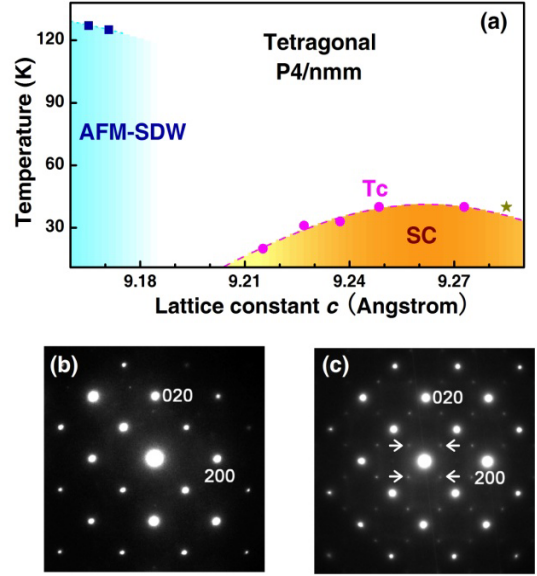

Figure 3. (a) Phase diagram of $(Li_{1-x}Fe_x)OHFeSe$ versus lattice parameter $c$, which is positively related to the doping level. Data represented by star symbol is taken from ref. [33]. (b) and (c) are the selected area diffraction patterns taken along the [001] zone axis direction. The former shows the basic tetragonal structure of the non-superconducting sample (SDW127), and the latter exhibits clearly visible satellite spots in the optimal superconducting sample (SC40-2), characterized by a unique modulation wave vector of $q = 1/2 (1, 1, 0)$. Such a superstructure dominates in the superconducting sample.

To conclude, a complete phase diagram of $(Li_{1-x}Fe_x)OHFeSe$ is obtained for the first time. Like the parent compounds of FeAs-based superconductors, the non-SC $(Li_{1-x}Fe_x)OHFeSe$ samples displays a AFM SDW order below 127 K in the absence of the well-known $\sqrt{5}\times\sqrt{5}$ ordered iron vacancies. Moreover, the unique superstructure, dominant in the SC phase with the modulation wave vector $q=1/2(1, 1, 0)$, and the comparable optimal $T_c$ of 40 K are very similar to previously reported for $K_yFe_{2-x}Se_2$ superconductors. Hence, we conclude that the high-$T_c$ superconductivity in $(Li_{1-x}Fe_x)OHFeSe$ emerges from the similarly weak AFM fluctuations, as it does in FeAs-based superconductors.[34] That strongly suggests that FeSe- and FeAs-based superconductors of the iron-based family may share a single superconductivity mechanism.

Further more, our results have indicated that superconductivity occurs only in the sufficiently intercalated $(Li_{1-x}Fe_x)OHFeSe$ samples with larger c-axis dimensions, where iron deficiency in the FeSe layers is small based on reported crystallographic data.

[31,32] These results may have a positive impact on the debate that the high-$T_c$ superconductivity with a comparable $T_c$ in $A_yFe_{2-x}Se_2$ system might not be generated within its so strong AFM background associated with the $\sqrt{5}\times\sqrt{5}$ ordered iron vacancies. Alternatively the SC phase in $A_yFe_{2-x}Se_2$ may be chemically, or spatially, phase-separated from the strong AFM phase. Maybe the unique superstructure in the SC phases of both $(Li_{1-x}Fe_x)OHFeSe$ and $K_yFe_{2-x}Se_2$ systems provides an important clue to resolve this issue. Further efforts are needed to find more new exemplifications, and the puzzles both in experiment and theory seem to converge towards a final answer.

**Corresponding Author**

xldong@iphy.ac.cn; zhxzhao@iphy.ac.cn

## ACKNOWLEDGMENT

X.L.D thanks Prof. Xianhui Chen from University of Science and Technology of China for his valuable help in hydrostatic synthesis technique. We thank Yue Wu, Mingwei Ma and Zeng Wang at IOP-CAS for their technical assistance. This work is supported by National Natural Science Foundation of China (projects 11274358 & 11190020), the National Basic Research Program of China (projects 2013CB921700 & 2011CB921703) and "Strategic Priority Research Program (B)" of the Chinese Academy of Sciences (No. XDB07020100 and XDB07020200).